\documentclass[preprint,prd,showpacs]{revtex4}
\newcommand{\bi}{\begin{itemize}}
\newcommand{\ei}{\end{itemize}}
\newcommand{\It}{\item}

\newcommand{\be}{\begin{equation}}
\newcommand{\ee}{\end{equation}}
\newcommand{\bdm}{\begin{displaymath}}
\newcommand{\edm}{\end{displaymath}}
\newcommand{\beqa}{\begin{eqnarray}}
\newcommand{\eeqa}{\end{eqnarray}}
\newcommand{\nonu}{\nonumber}
\newcommand{\des}{\partial\hspace{-4.mm}\not\hspace{2.5mm}}
\newcommand{\desco}{{\cal D}\hspace{-4.2mm}\not\hspace{2.2mm}}
\usepackage{epsfig}
\begin{document}
\title{Complete CKM quark mixing via Dimensional Deconstruction}
\author{P. Q. Hung }
\email{pqh@virginia.edu}
\affiliation{Institute for Nuclear and Particle Physics, and
Department of Physics, \\ University of Virginia, \\
Charlottesville, Virginia 22904-4714, USA}
\author{A. Soddu }
\email{asoddu@hep1.phys.ntu.edu.tw}
\affiliation{Department of Physics, National Taiwan University, \\
Taipei, Taiwan 106, R.O.C.}
\author{Ngoc-Khanh Tran }
\email{nt6b@virginia.edu}
\affiliation{Department of Physics, University of Virginia, \\
Charlottesville, Virginia 22904-4714, USA}
\date{\today}
\begin{abstract}
It is shown that the deconstruction of $[SU(2) \times U(1)]^N$
into $[SU(2) \times U(1)]$ is capable of providing all necessary
ingredients to completely implement the complex CKM mixing of
quark flavors. The hierarchical structure of quark masses originates
from the difference in the deconstructed chiral zero-mode
distributions in theory space, while the CP-violating phase
comes from the genuinely complex vacuum expectation value of link
fields. The mixing is constructed in a specific model to
satisfy experimental bounds on quarks' masses and CP violation.
\end{abstract}
 \maketitle
\section{Introduction}
Dimensional deconstruction \cite{ACG}, \cite{HPW} is a very
interesting approach to dynamically generate the effects of extra
dimensions departing from the four-dimensional (4D) renormalizable
physics at ultraviolet scale. That is, apart from having the
viability in the sense of renormalizability, whatever amusing
mechanisms being dynamically raised by the virtue of extra
dimensions (ED) now can also be easily arranged to rise
dynamically in a pure 4D framework. In this paper we look
specifically into two such important mechanisms of extra dimension
theories, namely the localization of matter fields in the bulk
\cite{JR,AS,HT} and the dynamical breaking of CP symmetry by ED
Wilson line \cite{CFH,CF,GW,Hosotani}. Ultimately, the hybrid of
these two mechanisms is just the well-known complex mixing of
fermion flavors. And it is conceptually interesting to note that
dimensional deconstruction (DD) nicely encompasses both of these
issues. In other words, complete Cabibbo-Kobayashi-Maskawa (CKM)
mixing can be generated naturally via dimensional deconstruction.

With the presence of extra dimensions, one has a new room to
localize the matter fields differently along the transverse
directions as in the so-called split fermion scenario. Various
overlaps of fermions of different flavors then induce various
fermion masses observed in nature (see e.g.
\cite{MirSchm,BGR,ST}). Amazingly, the deconstruction interaction
is also able to produce similar localization effects \cite{SS}.
Indeed, after the spontaneous breaking of link fields, fermions
get an extra contribution to their masses via the Higgs mechanism.
Fermions then reorganize themselves into mass sequences and the
lightest mass eigenstate of these towers exposes some interesting
``localization'' pattern in the theory space
(also referred to as deconstruction group index space). We will first
work out the analytical expressions and confirm the localization
of these zero modes in a rather generic deconstruction set-up. The
next question to raise is how to make these light modes chiral.
Imposing some kind of chiral boundary conditions \cite{HPW} is the
answer again coming from the ED lessons. There is however one more
subtle point to be mentioned here. If one truly wishes to relate
the ED scenario to the dimensional deconstruction, one needs to
latticize the extra dimensions to host the deconstruction group.
There comes the lattice theory's issue of fermion doubling, and
its standard remedy, such as adding to the Lagrangian a Wilson
term \cite{HL} would remove half of original chiral degrees of
freedom. This is the reason why most of previous works addressing
the fermionic
mixing in deconstructed picture (e.g. \cite{HPW,SS,AKMY}) usually
start out with only Weyl spinors. In the current work, we adopt a
different and somewhat more general 4D deconstruction approach
\cite{HTn} where no extra dimension is actually invoked. As a
result the fermions to begin with keep a standard 4-component
Dirac spinor representation.

In any deconstruction set-up, the link fields transform
non-trivially under at least two different gauge groups. This
implies a complex vacuum expectation value (VEV) for these fields,
whose phase would not be rotated away in general. After the
deconstruction process, this phase is carried over into the
complex value of  wave functions and wave function overlaps of
fermions. In turn, the induced complex-valued mass matrices can
render a required CP-violating phase in the well-known KM
mechanism. In contrast, we note that the generation of complex
mass matrices within the split fermion scenario is a non-trivial
problem and requires rather sophisticated techniques to solve
\cite{HS,HSS}. Interestingly, the above CP violation induction via
deconstruction can also be visualized in extra dimensional view
point. Indeed, because of having the same symmetry transformation
property, DD link field can be identified with the Wilson line
pointing along a latticized transverse direction (Appendix B), and
the latter then can naturally acquire a complex VEV in the
generalized Hosotani's mechanism \cite{Hosotani, CFH, CF, GW} of
dynamical symmetry breaking. Apparently, the source of CP
violation in this approach comes from the complex effective Yukawa
couplings so it can be classified as hard CP violation.
Nevertheless, those couplings acquire complex values after the
spontaneous breaking of the DD link fields. In that sense this CP
violation pattern could also be considered soft and dynamical.

This paper is presented in the following order. In Sec. IIA we
give the zero mass eigenstate of fermions obtained in the
deconstructed picture, in Sec. IIB the resulting expression of
mass matrix elements, and in Sec. IIC the symmetry
breaking of $[SU(2) \times U(1)]^N$ into $[SU(2) \times U(1)]$.
In Sec. III we present the numerical fit for
quark mass spectrum and CKM matrix in a model where each ``standard
model'' Higgs field is chosen to transform under only a single
deconstruction subgroup. The conclusion and
comments on numerical results is given in Sec. V. Appendix A provides
a detailed derivation of zero mode wave functions in 4D deconstruction using
combinatoric techniques. Appendix B outlines intuitive arguments
on the complexity of link field inspired by lattice models.
Appendix C presents analytical expressions for wave function
overlaps used in the determination of mass matrix elements.
Finally, Appendix D gives referencing values of key physical
quantities that have been used in the search algorithm (Table I),
and numerical solution of our models' parameters (Table II).
\section{Deconstruction and Quark Mass Matrix}
In this section we describe how the mixing of quark flavors arises
in the DD picture. But we first briefly recall the basic idea of
the dimensional deconstruction applied to just a single quark
generation. The family replication will be restored in the later
sections.
\subsection{Zero-mode fermion}
We begin with N copies of gauge group $[SU(2)\times U(1)]_n$ where
$n=1,..,N$. To each group $[SU(2)\times U(1)]_n$ we associate a
$SU(2)_n$-doublet $Q_n$, and two $SU(2)_n$-singlets $U_n$, $D_n$.
These fields transform non-trivially only under their
corresponding group $[SU(2)\times U(1)]_n$ as $(2,q_Q)$,
$(1,q_U)$, $(1,q_D)$ respectively, with $q$'s denoting $U(1)$-charges.
Finally, we use $3(N-1)$
scalars $\phi^{Q}_{n-1,n}$, $\phi^{U}_{n-1,n}$, $\phi^{D}_{n-1,n}$
transforming respectively as $(2,q_Q|2,-q_Q)$,
$(1,q_U|1,-q_U)$, $(1,q_D|1,-q_D)$ under $[SU(2)\times U(1)]_{n-1}
\times [SU(2)\times U(1)]_n$ to ``link'' fermions of the same
type. Because of this, scalars $\phi$'s are also referred to as
link fields hereafter. For the simplicity of the model, we
assume a symmetry for the Lagrangian under the permutation of
group index $n$.

The $\prod_{n=1}^N [SU(2) \times U(1)]_n$ gauge-invariant
Lagrangian of the fermionic sector is \be \label{L}
{\cal{L}}=\left(\sum_{n=1}^{N} \bar{Q}_n i\desco_n Q_n +
\sum_{n=1}^{N-1} \bar{Q}_n \phi^{Q}_{n,n+1} Q_{n+1} -
M_Q\sum_{n=1}^{N} \bar{Q}_n  Q_n\right) +\left( Q \leftrightarrow
U\right) + \left(Q \leftrightarrow D \right) \, ,\ee where
$\desco_n$ denotes the covariant derivative associated with gauge
group $[SU(2)\times U(1)]_n$, and $M_{Q},M_{U},M_{D}$ are the bare
masses of fermions. Ultimately, we are interested in achieving
chiral fermions of standard model (SM) at low energy scale. To
this aim we impose the following chiral boundary conditions (CBC)
on fermion fields \cite{HPW}.
\beqa \nonu Q_{1R}=Q_{NR}=0\, ,
\;\;\;\;\;\;\;\;\;\;
\phi^Q_{N-1,N} Q_{N,L} =V_Q Q_{N-1,L} \, , \\
\label{CBC} U_{1L}=U_{NL}=0\, , \;\;\;\;\;\;\;\;\;\;
\phi^U_{N-1,N} U_{N,R} =V_U U_{N-1,R} \, ,\\ \nonu
D_{1L}=D_{NL}=0\, , \;\;\;\;\;\;\;\;\;\; \phi^D_{N-1,N} D_{N,R}
=V_D D_{N-1,R} \, . \eeqa We note that those conditions are in
agreement with the gauge transformation property of fields, e.g.
$\phi^Q_{N-1,N} Q_{N,L}$ and $Q_{N-1,L}$ transform identically
under the underlying gauge groups. Essentially, these boundary
conditions render one more left-handed degree of freedom over the
right-handed for $Q$ field, and the contrary holds for $U$ and $D$
fields. The actual calculation will show that the zero-mode of $Q$
field indeed is left-handed while for $U,D$ it is right-handed.
When the link fields $\phi^{Q,U,D}$ assume VEV proportional to
$V_{Q,U,D}$, above CBC become the very reminiscence of Neumann and
Dirichlet boundary conditions.

In the deconstruction scenario, after the spontaneous
symmetry breaking (SSB) the link fields acquire an uniform VEV
$V_{Q,U,D}$ respectively, independent of site index $n$ (in accordance
with the assumed permutation symmetry), and the
fermions obtain new mass structure. Using the CBC (\ref{CBC}), the
fermion mass term can be written in the chiral basis as \beqa
\nonu (\bar{Q}_{2R},..,\bar{Q}_{N-1,R})\; [M_Q] \; \left(
\begin{array}{c}
Q_{1L} \\
: \\
Q_{N-1,L}
\end{array} \right)
\;\;&+&\;\; (\bar{Q}_{1L},..,\bar{Q}_{N-1,L})\;[M_Q]^{\dagger}
\left( \begin{array}{c}
Q_{2R} \\
: \\
Q_{N-1,R}
\end{array} \right) \\ +
\left( Q_{R,L} \leftrightarrow U_{L,R}\right) \;\;&+&\;\;
\left(Q_{R,L} \leftrightarrow D_{L,R} \right) \, ,\eeqa where the
matrix $[M_Q]$ of dimension $(N-2)\times (N-1)$ is \be
[M_Q]_{(N-2)\times (N-1)}= \left( \begin{array}{ccccccc}
-V_Q^{*}&M_Q&-V_Q&0& & &   \\
0&-V_Q^{*}&M_Q&-V_Q& & &   \\
0& 0&-V_Q^{*}&M_Q  & & &   \\
&&&&\ddots&& \\
&&&&\;\;\;\;&M_Q&-V_Q \\
&&&&\;\;\;\;&-V_Q^{*}& M_Q - V_Q
\end{array} \right) \, .
\ee By interchanging $Q_{R,L} \leftrightarrow U_{L,R}$, $Q_{R,L}
\leftrightarrow D_{L,R}$, the matrices $[M_U]$, $[M_D]$ of
dimension $(N-1)\times (N-2)$ can be analogously found.

By coupling the following Dirac equations for chiral fermion sets
$\{ Q_R \} \equiv (Q_{2R},..,Q_{N-1,R})^T$ and $\{ Q_L \} \equiv
(Q_{1L},..,Q_{N-1,L})^T$ \be i\des\{ Q_R \} - [M_Q] \{Q_L\} = 0 \,
, \;\;\;\;\;\;\;\;\; i\des\{ Q_L \} - [M_Q]^{\dagger} \{Q_R\} = 0
\, , \ee we see that $[M_Q^{\dagger}M_Q]$ is the squared-mass
matrix for the left-handed components $Q_L$ and $[M_Q
M_Q^{\dagger}]$ for the right-handed $Q_R$. Since at low energy,
we are interested only in the chiral zero modes of fermions, we
will work only with $[M_Q^{\dagger}M_Q]$, $[M_U M_U^{\dagger}]$,
$[M_D M_D^{\dagger}]$ in what follows. {\scriptsize \beqa
[M_Q^{\dagger} M_Q]&=& \left(
\begin{array}{ccccccccc}
|V_Q|^2 & -M_Q V_Q & V_Q^2 & 0   & & & & &  \\
-M_QV_Q^{*} &
\begin{array}{c} M_Q^2 +\\|V_Q|^2 \end{array}
& -2M_Q V_Q & V_Q^2 &  & & & &   \\
(V_Q^{*})^2 & -2M_Q V_Q^{*} &
\begin{array}{c} M_Q^2 +\\2|V_Q|^2 \end{array}
& -2M_Q V_Q  &  & & & & \\
0 & (V_Q^{*})^2 & -2M_Q V_Q^{*} &
\begin{array}{c} M_Q^2 +\\2|V_Q|^2 \end{array}
&  & & & & \\
&&&&\ddots&&&& \\
&&&&& -2M_Q V_Q^{*} &
\begin{array}{c} M_Q^2 +\\2|V_Q|^2 \end{array}
& -2M_Q V_Q  & V_Q^2 \\
&&&&& (V_Q^{*})^2 & -2M_Q V_Q^{*} &
\begin{array}{c} M_Q^2 +\\2|V_Q|^2 \end{array}
&
\begin{array}{c} -2M_Q V_Q +\\|V_Q|^2 \end{array}
\\
&&&&& 0 & (V_Q^{*})^2 &
\begin{array}{c} -2M_Q V_Q^{*} +\\ (V_Q^{*})^2 \end{array}
&\begin{array}{c} M_Q^2 + 2|V_Q|^2 - \\ (V_Q+V_Q^{*})M_Q
\end{array}
\end{array} \right) \, ,\nonumber
\eeqa }
\begin{equation}
\label{M^2_Q}
\end{equation}
and similar expressions hold for $[M_U M_U^{\dagger}]$,
$[M_D M_D^{\dagger}]$. The quantitative derivation of the
zero-eigenstates, which are identified with the SM chiral fermion,
is presented in appendix A. In this section we just concentrate on
some qualitative discussion. In general the diagonalization of
matrices (\ref{M^2_Q}) leads to the transformation between gauge
eigenstates $Q_{nL}$ and mass eigenstates $\tilde{Q}_{mL}$ \be
\label{MassState} Q_{nL} = [{\cal U}_Q]_{nm} \tilde{Q}_{mL} \, ,
\;\;\;\;\;\;\; \tilde{Q}_{nL} = [{\cal U}_Q]^{*}_{mn} Q_{mL} \,
,\ee where the matrix $[{\cal{U}}_Q]$ diagonalizes
$[M_Q^{\dagger}M_Q]$ \be \label{dial} [M_Q^{\dagger}M_Q]_{diag} =
[{\cal{U}}_Q]^{\dagger}[M_Q^{\dagger}M_Q][{\cal{U}}_Q] \, .\ee The
key observation, which will be analyzed in more details in
appendix B, is that VEV $V_{Q,U,D}$ are generically complex and
$[\cal{U}_{Q,U,D}]$ are truly unitary (i.e. not just orthogonal).
This in turn gives non-trivial phases to zero-mode fermions
$\tilde{Q}_{0L}, \tilde{U}_{0L}, \tilde{D}_{0L}$ in Eq.
(\ref{MassState}) and after the SM spontaneous symmetry breaking
the obtained mass matrices are complex. Further, the explicit
solution of zero mode $\tilde{Q}_{0L}$ (and $\tilde{U}_{0R}$,
$\tilde{D}_{0R}$) in the mass eigenbasis exhibits a very
interesting ``localization'' pattern in the group index space n
(see Appendix A). This in turn can serve to generate the mass
hierarchy among fermion families in a manner similar to that of ED
split fermion scenario (see e.g. \cite{HSS,ST}). Thus we see that
dimensional deconstruction indeed provides all necessary
ingredients to construct a complete (complex) CKM structure of
fermion family mixing.
\subsection{Complex Mass Matrix}
In order to give mass to the above chiral zero-mode of fermions,
we introduce Higgs doublet fields just as in the SM. In the
simplest and most evident scenario (see \cite{AKMY}), there is one
doublet Higgs $H_n$ transforming as $(2,q_Q-q_D\equiv q_U-q_Q)$
under each $[SU(2)\times U(1)]_n$ group. We also implement the
replication of families by incorporating family indices $i,j
=1,..,3$. Another scenario to generate the (vector-like) fermion mass
hierarchy by assuming various link fields to connect arbitrary sites of
the latticized fifth dimension has been proposed in \cite{NOS}.

The gauge-invariant Yukawa terms read \be \label{Ygauge}
\kappa^U_{ij} \sum_{n=1}^{N} \bar{Q}^{(i)}_n i\sigma_2 H_n^{*}
U^{(j)}_n + \kappa^D_{ij} \sum_{n=1}^{N} \bar{Q}^{(i)}_n  H_n
D^{(j)}_n + H.c. \, .\ee In order to extract the terms involving
zero modes, which are the only terms relevant at low energy limit,
we rewrite (\ref{Ygauge}) in the mass eigenbasis. However, this
procedure depends explicitly on the specific CBCs being imposed on
each of the fields Q, U, D. To be generic, let us consider the
following configuration. We assume the ``localization'' of zero
modes $\tilde{Q}_{0L}$, $\tilde{U}_{0R}$ and $\tilde{D}_{0R}$ to
be at $n=1$, $n=1$ and $n=N$ respectively. To achieve this
localization pattern, we impose the following CBCs on these fields
(see Eq. (\ref{CBC}) and Appendix A, Eq. (\ref{yCBC})) \beqa \nonu
&&Q^{(i)}_{1R}=Q^{(i)}_{NR}=0\, , \;\;\;\;\;\;\;\;\;\;
\phi^{(i)Q}_{N-1,N} Q^{(i)}_{NL} =V^{(i)}_Q Q^{(i)}_{N-1L} \, ,\\
\label{qudCBC} &&U^{(j)}_{1L}=U^{(j)}_{NL}=0\, ,
\;\;\;\;\;\;\;\;\;\;
\phi^{(j)U}_{N-1,N} U^{(j)}_{NR} =V^{(j)}_U U^{(j)}_{N-1R} \, , \\
\nonu &&D^{(k)}_{1L}=D^{(k)}_{NL}=0 \, , \;\;\;\;\;\;\;\;\;\;
{\phi^{(k)D}_{1,2}}^{\dagger} D^{(k)}_{1R} = V^{(k)*}_D
D^{(k)}_{2R} \, .\eeqa Because of these boundary conditions, zero
modes $\tilde{Q}_{0L}$, $\tilde{U}_{0R}$ and $\tilde{D}_{0R}$
would be localized at $n=1$, $n=1$ and $n=N$ respectively, this
also means that the first term of Eq. (\ref{Ygauge}) would
represent the overlap between 2 wave function localized at the same
site $n=1$, while the second term represents the overlap between
wave functions localized at $n=1$ and $n=N$. Using (\ref{qudCBC})
to eliminate the dependent components and after the SM spontaneous
symmetry breaking $\langle H_n\rangle = (0,v/\sqrt{2})^T$
uniformly for all $n$'s, we can rewrite the Yukawa term
(\ref{Ygauge}) as \beqa \nonu & &\kappa^U_{ij}\frac{v }{\sqrt{2}}
\sum_{i,j=1}^{3}\left[\bar{Q}^{(i)}_{1L}U^{(j)}_{1R} +
\frac{V^{(i)}_Q}{\phi^{(i)Q}_{N-1,N}}
\frac{V^{(j)}_U}{\phi^{(j)U}_{N-1,N}} \bar{Q}^{(i)}_{N-1L}
U^{(j)}_{N-1R} + \sum_{n=2}^{N-1} (\bar{Q}^{(i)}_{nL} U^{(j)}_{nR}
+ \bar{Q}^{(i)}_{nR} U^{(j)}_{nL})\right] +
\\
& & \kappa^D_{ij}\frac{ v}{\sqrt{2}} \sum_{i,k=1}^{3}
\left[\frac{V^{(k)*}_D}{{\phi^{(k)D}_{1,2}}^{\dagger}}
\bar{Q}^{(i)}_{1L}D^{(k)}_{2R} +
\frac{V^{(i)}_Q}{\phi^{(i)Q}_{N-1,N}} \bar{Q}^{(i)}_{N-1L}
D^{(k)}_{NR} + \sum_{n=2}^{N-1} (\bar{Q}^{(i)}_{nL} D^{(k)}_{nR} +
\bar{Q}^{(i)}_{nR} D^{(k)}_{nL})\right ] \, .\eeqa After going to
the mass eigenbasis by the virtue of transformation of the type
(\ref{MassState}), keeping only zero-mode terms and together with
the assumption of universality for the Yukawa couplings in the up
and down sectors, we obtain the following effective mass terms \be
\sum_{i,j=1}^{3} \bar{\tilde{Q}}^{(i)}_{0L} M^u_{ij}
\tilde{U}^{(j)}_{0R} +\sum_{i,k=1}^{3} \bar{\tilde{Q}}^{(i)}_{0L}
M^d_{ik} \tilde{D}^{(k)}_{0R} \, ,\ee with \beqa \label{MU}
M^u_{ij} =&&\kappa_U\frac{v}{\sqrt{2}}\left[\left(
\sum_{n=1}^{N-2} [{\cal{U}}_Q^{(i)}]^{*}_{n,0}
[{\cal{U}}_U^{(j)}]_{n,0} \right)
+\left(1+\frac{V^{(i)}_Q}{\phi^{(i)Q}_{N-1,N}}
\frac{V^{(j)}_U}{\phi^{(j)U}_{N-1,N}}\right)
[{\cal{U}}_Q^{(i)}]^{*}_{N-1,0} [{\cal{U}}_U^{(j)}]_{N-1,0}
\right] \, ,\nonumber
\\ \\
\label{MD} M^d_{ik} = && \kappa_D\frac{v}{\sqrt{2}}\left[\left(
\frac{V^{(k)*}_D}{{\phi^{(k)D}_{1,2}}^{\dagger}}
[{\cal{U}}_Q^{(i)}]^{*}_{1,0} +[{\cal{U}}_Q^{(i)}]^{*}_{2,0}
\right) [{\cal{U}}_D^{(k)}]_{2,0} \right.\nonumber  \\ + &&
\left.\left( \sum_{n=3}^{N-2} [{\cal{U}}_Q^{(i)}]^{*}_{n,0}
[{\cal{U}}_D^{(k)}]_{n,0} \right) +[{\cal{U}}_Q^{(i)}]^{*}_{N-1,0}
\left([{\cal{U}}_D^{(k)}]_{N-1,0} +
\frac{V^{(i)}_Q}{\phi^{(i)Q}_{N-1,N}} [{\cal{U}}_D^{(k)}]_{N,0}
\right)\right] \, .\eeqa Because all $[{\cal{U}}_Q],
[{\cal{U}}_U], [{\cal{U}}_D]$ are unitary, the mass matrices
$M^u$, $M^d$ are generally complex. Thus in this simplest
deconstruction approach, we might better understand the dynamical
origin of CP-violation phase in the SM mass matrices. We also note
that (\ref{MU}), (\ref{MD}) represent the specific case where
$\tilde{Q}_{0L}$, $\tilde{U}_{0R}$ and $\tilde{D}_{0R}$ are
localized at $n=1$, $n=1$ and $n=N$ respectively. All other
localization configurations can be similarly found. Further, when
we replace link fields $\phi$'s in (\ref{MU}), (\ref{MD}) by their
VEVs following the deconstruction, these mass matrix elements will
look much simpler (see (\ref{overlapU}), (\ref{overlapD})).

Before moving on to give explicit expressions of these
complex-valued mass matrices in term of zero mode wave functions
(appendix A) and perform the numerical fit, let us briefly turn to the
breaking pattern of product group
$\prod_{n=1}^N [SU(2) \times U(1)]_n$.
\subsection{Deconstructing $[SU(2) \times U(1)]^N$}
For the sake of completeness, in this section we will describe
the breaking of $[SU(2) \times U(1)]^N$ into the SM
$[SU(2) \times U(1)]$ gauge group by giving uniform VEVs to link
fields. The transformation and charge structure of fermions and
scalar link fields have been defined in the beginning of previous
section. To identify the unbroken symmetries following the deconstruction,
we look at the covariant derivative and kinetic terms of scalars.
\beqa
\label{DU}
D_{\mu} \phi^U_{n,n+1}=\partial_{\mu} \phi^U_{n,n+1} -
iq_U\frac{g'_0}{2}B_{n\mu} \phi^U_{n,n+1}+
iq_U\frac{g'_0}{2}B_{n+1\mu} \phi^U_{n,n+1}\, , \\
\label{DD}
D_{\mu} \phi^D_{n,n+1}=\partial_{\mu} \phi^D_{n,n+1} -
iq_D\frac{g'_0}{2}B_{n\mu} \phi^D_{n,n+1}+
iq_D\frac{g'_0}{2}B_{n+1\mu} \phi^D_{n,n+1}\, .
\eeqa
where $B_n$ is the gauge boson associated with $U(1)_n$, while
$g'_0$ is the common gauge coupling for all $U(1)$'s. For abelian
groups, the opposite signs of the last two terms in (\ref{DU})
(and also in (\ref{DD})) originate from the opposite charges of
$\phi^U_{n,n+1}$
(and $\phi^D_{n,n+1}$) under $U(1)_n$ and $U(1)_{n+1}$ (so that terms like
$\bar{U}_n \phi^U_{n,n+1}U_n$ are gauge-invariant).

For non-abelian groups, the similar sign reversing will hold for terms in the
expression of covariant derivatives (see Eq. (\ref{DQ}) below),
the nature of which also has its
root in the gauge invariance of the theory. Indeed, under the
Yang-Mills $SU(2)_n \times SU(2)_{n+1}$ gauge tranfromation
(note that $\phi^Q_{n,n+1}$ is a $2\times 2$-matrix)
\beqa
\label{TPhi}
&&\phi^Q_{n,n+1} \rightarrow  T_n  \phi^Q_{n,n+1} T_{n+1}^{\dagger}\, , \\
\label{TQ}
&&Q_n \rightarrow  T_n Q_n\, , \;\;\;\;\;\;\;\;\;\;
Q_{n+1} \rightarrow  T_{n+1} Q_{n+1}\, , \\
\label{TA1}
&&\left[\vec{A}_{n\mu}\frac{\vec{\tau}}{2} \right] \rightarrow
T_n \left[\vec{A}_{n\mu}\frac{\vec{\tau}}{2} \right] T_n^{\dagger}
- \frac{i}{g_0} (\partial_{\mu} T_n) T_n^{\dagger}\, , \\
\label{TA2} &&\left[\vec{A}_{n+1\mu}\frac{\vec{\tau}}{2} \right]
\rightarrow T_{n+1}\left[\vec{A}_{n+1\mu}\frac{\vec{\tau}}{2}
\right] T_{n+1}^{\dagger} - \frac{i}{g_0} (\partial_{\mu} T_{n+1})
T_{n+1}^{\dagger}\, . \eeqa the covariant derivative of
$\phi^Q_{n,n+1}$ must be formulated as follows (so that it
transforms exactly like $\phi^Q_{n,n+1}$ in (\ref{TPhi})) \beqa
\nonu D_{\mu} \phi^Q_{n,n+1}&=&\partial_{\mu} \phi^Q_{n,n+1} -
(iq_Q\frac{g'_0}{2}B_{n\mu} \phi^Q_{n,n+1}+
ig_0 \vec{A}_{n\mu}\frac{\vec{\tau}}{2} \phi^Q_{n,n+1}) \\
&&  +(iq_Q\frac{g'_0}{2}B_{n+1\mu} \phi^Q_{n,n+1}+ ig_0
\phi^Q_{n,n+1}\vec{A}_{n\mu}\frac{\vec{\tau}}{2} )\, . \label{DQ}
\eeqa where $\vec{A}_{n}$ and $T_n$ are respectively the gauge bosons
and some $2\times 2$-special unitary matrix characterizing the
$SU(2)_n$ transformation, while $g_0$ is the common gauge coupling
for all $SU(2)$'s.

After the deconstruction $\phi^{U,D}_{n,n+1}\rightarrow V_{U,D}$,
$\phi^{Q}_{n,n+1}\rightarrow V_{Q} \cdot {\bf 1}_{2\times 2}$, the
mass terms for gauge bosons are generated. Specifically, we obtain
as parts of kinetic terms $(D_{\mu}
\phi^U_{n,n+1})^{\dagger}(D^{\mu} \phi^U_{n,n+1})$, $(D_{\mu}
\phi^D_{n,n+1})^{\dagger}(D^{\mu} \phi^D_{n,n+1})$, Tr[$(D_{\mu}
\phi^Q_{n,n+1})^{\dagger}(D^{\mu} \phi^Q_{n,n+1})$] the following
gauge bosons squared mass matrices \be \label{MBA} [M_B^2]=
\lambda_B \left( \begin{array}{ccccc}
1&-1&&& \\
-1&2&&& \\
&&\ddots&& \\
&&&2&-1\\
&&&-1&1 \end{array} \right);
\;\;\;\;\;\;\;
[M_{\vec{A}}^2]= \lambda_{\vec{A}} \left( \begin{array}{ccccc}
1&-1&&& \\
-1&2&&& \\
&&\ddots&& \\
&&&2&-1\\
&&&-1&1 \end{array}\right) \, .
\ee
where, after restoring the family replication index ($i=1,2,3$),
\be
\label{lambda}
\lambda_B = \sum_{1}^{3} {g_0'}^2 (q_U^2 |V^{(i)}_U|^2 + q_D^2 |V^{(i)}_D|^2
+ q_Q^2 |V^{(i)}_Q|^2  )\, , \;\;\;\;\;
\lambda_{\vec{A}} = \sum_{1}^{3} g_0^2 |V^{(i)}_Q|^2   \, .
\ee
Both matrices in (\ref{MBA}) have a ``flat'' zero eigenstate. This
indeed indicates the uniform breaking of $[SU(2)\times U(1)]^N$ into
the diagonal (SM) group $[SU(2)\times U(1)]$, whose gauge bosons are
massless and given by
\be
\label{BA}
\tilde{B}_{\mu}= \frac{1}{\sqrt{N}} \sum_{n=1}^{N} B_{n\mu}\, ,
\;\;\;\;\;\;\;\;\;\;
\vec{{\tilde{A}}}_{\mu}= \frac{1}{\sqrt{N}} \sum_{n=1}^{N} \vec{A}_{n\mu} \, .
\ee
In Eq. (\ref{lambda}) it is also shown that the pattern of symmetry breaking
is not spoiled by family replication as long as charges $q_U$
(and $q_D$, $q_Q$ ) are independent of the site index $n$ under a presumed
permutation symmetry (just like $V^{(i)}_{U,D,Q}$). Finally, by extracting the
interaction between fermions and massless gauge bosons from fermion kinetic
terms in (\ref{L}) one can see that the couplings of the unbroken
group scale as $g'=g'_0/\sqrt{N}$ and $g=g_0/\sqrt{N}$,
while the charge structure (of fermions in mass eigenbasis) under this
diagonal group remains intact.
\section{Fitting the Model's Parameters}
\subsection{Model, Parameters and Numerical Method}
In the previous section we have outlined the process diagonalizing
the squared-mass matrix (\ref{M^2_Q}). The complete
diagonalization process is complicated, but as we are concerned
only with the zero eigenvalue problem, the computation can be done
analytically in the general term (see Appendix A). Since
$[{\cal{U}}_Q]$ diagonalizes $[M_Q^{\dagger}M_Q]$ (\ref{dial}),
the zero eigenstate of $[M_Q^{\dagger}M_Q]$ is just the first
column of $[{\cal{U}}_Q]$, i.e. in the notation of appendix A \be
\label{Ux} [{\cal{U}}_Q^{(i)}]_{n,0} = x_{Qn}^{(i)} \, ,\ee and
similarly \be \label{Uy} [{\cal{U}}_U^{(j)}]_{n,0} =
x_{Un}^{(j)}\, , \;\;\;\;\;\;\;\;\;\; [{\cal{U}}_D^{(k)}]_{n,0} =
y_{Dn}^{(k)} \, ,\ee where $x_n$'s are given in (\ref{Newx_ns})
(corresponding to a zero mode localized at the end point $n=1$)
and $y_n$'s in (\ref{y_ns}) (corresponding to a zero mode
localized at the end point $n=N$).

After the spontaneous symmetry breaking, the link fields acquire
an uniform VEV $V_{Q,U,D}$ respectively (independent of site index
$n$). In term of $x_n$'s and $y_n$'s, the SM mass matrices
(\ref{MU}), (\ref{MD}) for up and down quark sectors become \beqa
\label{overlapU} M^u_{ij} =&&
\kappa_U\frac{v}{\sqrt{2}}\left[\left( \sum_{n=1}^{N-1}
x_{Qn}^{(i)*} x_{Un}^{(j)} \right) + x_{QN-1}^{(i)*}x_{UN-1}^{(j)}
\right] \, ,
\\
\label{overlapD} M^d_{ik} =&&
\kappa_D\frac{v}{\sqrt{2}}\left[x_{Q1}^{(i)*} y_{D2}^{(k)} +
\left(\sum_{n=2}^{N-1} x_{Qn}^{(i)*}y_{Dn}^{(k)} \right) +
x_{QN-1}^{(i)*} y_{DN}^{(k)} \right] \, ,\eeqa where $x_n$'s,
$y_n$'s are given in (\ref{Newx_ns}), (\ref{y_ns}) respectively.
The analytical forms of (\ref{overlapU}), (\ref{overlapD}) in term
of model's parameters are worked out in Appendix C, Eqs. (\ref{xx20}),
(\ref{xy20}).

Again, let us remind ourselves that (\ref{overlapU}) represents
the overlap between two wave functions localized at the same site
$n=1$ while (\ref{overlapD}) represents the overlap between one
wave function localized at $n=1$ and the other at $n=N$. The model
under consideration consists of 20 real parameters (see Table II
and appendix B): 3 complex VEV $V$'s for each complete quark
generation $(Q,U,D)_i$ (i=1,2 or 3), and 2 real ``dimensionful''
Yukawa couplings  $\kappa_U v/\sqrt{2}$,
$\kappa_D v/\sqrt{2}$. We choose to fix $N=10$ throughout.

The numerical approach to fit the parameters consists in
minimizing a positive function which gets a zero value when all
the predicted quantities are in the corresponding experimental
ranges \cite{HSS}. The minimization procedure is based on the
simulated annealing method, which seems working better than other
minimization approaches when the parameter space becomes larger
\cite{SA1},\cite{NR}. The input referencing physical quantities
are given in Table I of Appendix D.

We consider eight different cases, which correspond to all the
eight possible ways of localizing the left and right components.
The eight different cases are the following:

\begin{enumerate}
\item{$Q$, $U$ and $D$ localized in $n=1$ denoted as $(QUD1)$}
\item{$Q$ and $U$ localized in $n=1$,  $D$ localized in $n=N$
denoted as $(QU1DN)$}
\item{$Q$ and $D$ localized in $n=1$, $U$
localized in $n=N$ denoted as $(QD1UN)$}
\item{$Q$ localized in
$n=1$, $U$ and $D$ localized in $n=N$ denoted as $(Q1UDN)$}
\item{$Q$, $U$ and $D$ localized in $n=N$ denoted as $(QUDN)$}
\item{$D$ localized in $n=1$, $Q$ and $U$ localized in $n=N$
denoted as $(D1QUN)$}
\item{$U$ localized in $n=1$, $Q$ and $D$
localized in $n=N$ denoted as $(U1QDN)$}
\item{$U$ and $D$
localized in $n=1$, $Q$ localized in $n=N$ denoted as $(UD1QN)$}
\end{enumerate}

We specially note that, due to the mirror complexity between CBCs
(\ref{CBC}) and  (\ref{yCBC}), the mass matrices obtained in the
cases 1 and 5, cases 2 and 6, cases 3 and 7, cases 4 and 8, are
complex conjugate pairwise. In the result, all eight cases are
inequivalent.

\subsection{Numerical results}

\begin{figure}
        \begin{center}
        \epsfig{figure=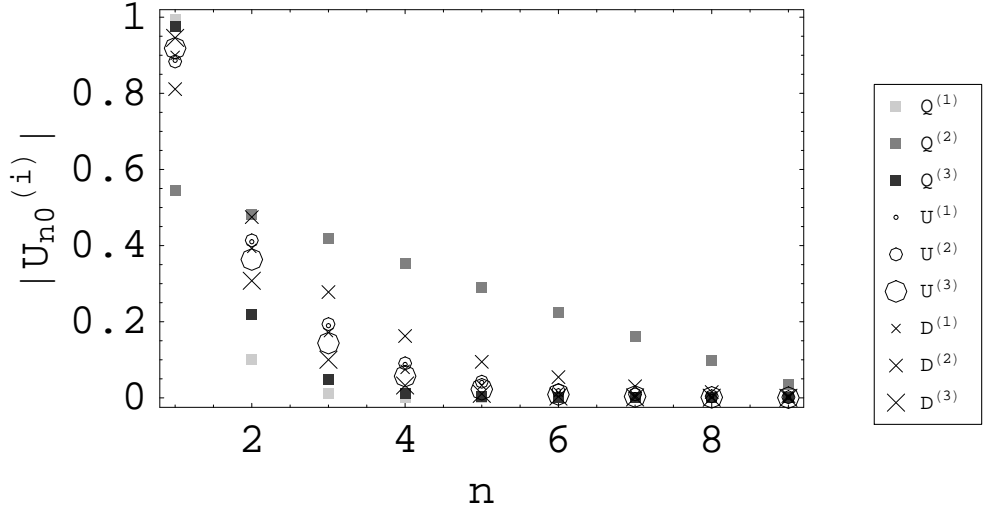,width=0.85\textwidth}
        \end{center}
        \vspace{0cm}
        \caption{Profiles of the absolute value of normalized wave functions
$|x_{Qn}^{(i)}|$, $|x_{Un}^{(i)}|$ and $|x_{Dn}^{(i)}|$ in the
theory space (N=10) for the case with $Q$, $U$ and
$D$ localized at n=1. $|x_{Qn}^{(2)}|$ with a value of $\alpha \ll 1$ is less localized. }
        \label{Figure_1}
        \end{figure}

\begin{figure}
        \begin{center}
        \epsfig{figure=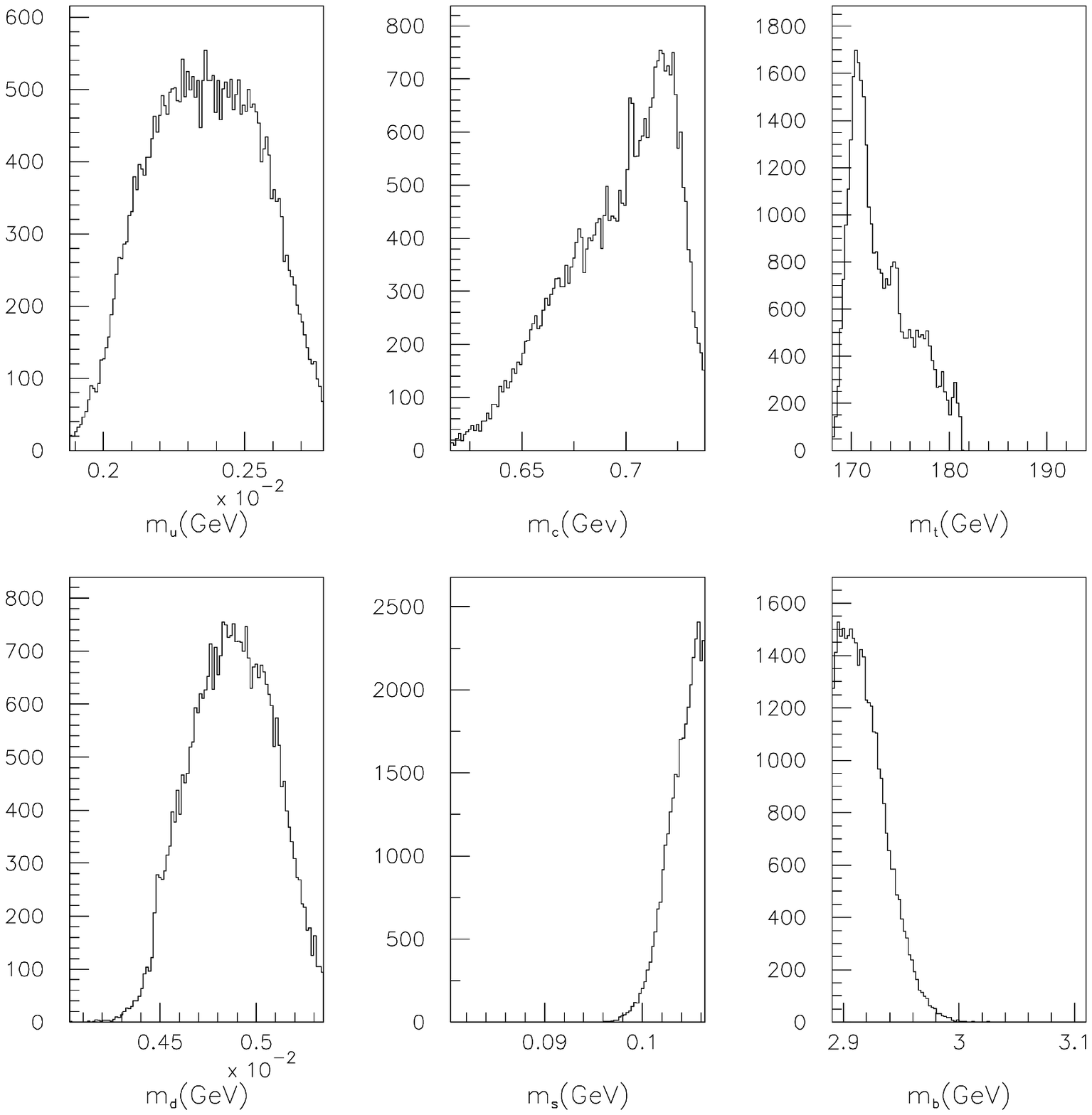,width=0.65\textwidth}
        \end{center}
        \vspace{0cm}
        \caption{Solutions for the 6 quark masses in the 
              case with $Q$, $U$ and $D$ localized at n=1. The
              masses in $GeV$ are evaluated at the $M_Z$ scale. The range for
                 each mass is given by the edges of the corresponding window.
        \label{Figure_2}}
        \end{figure}

\begin{figure}
        \begin{center}
        \epsfig{figure=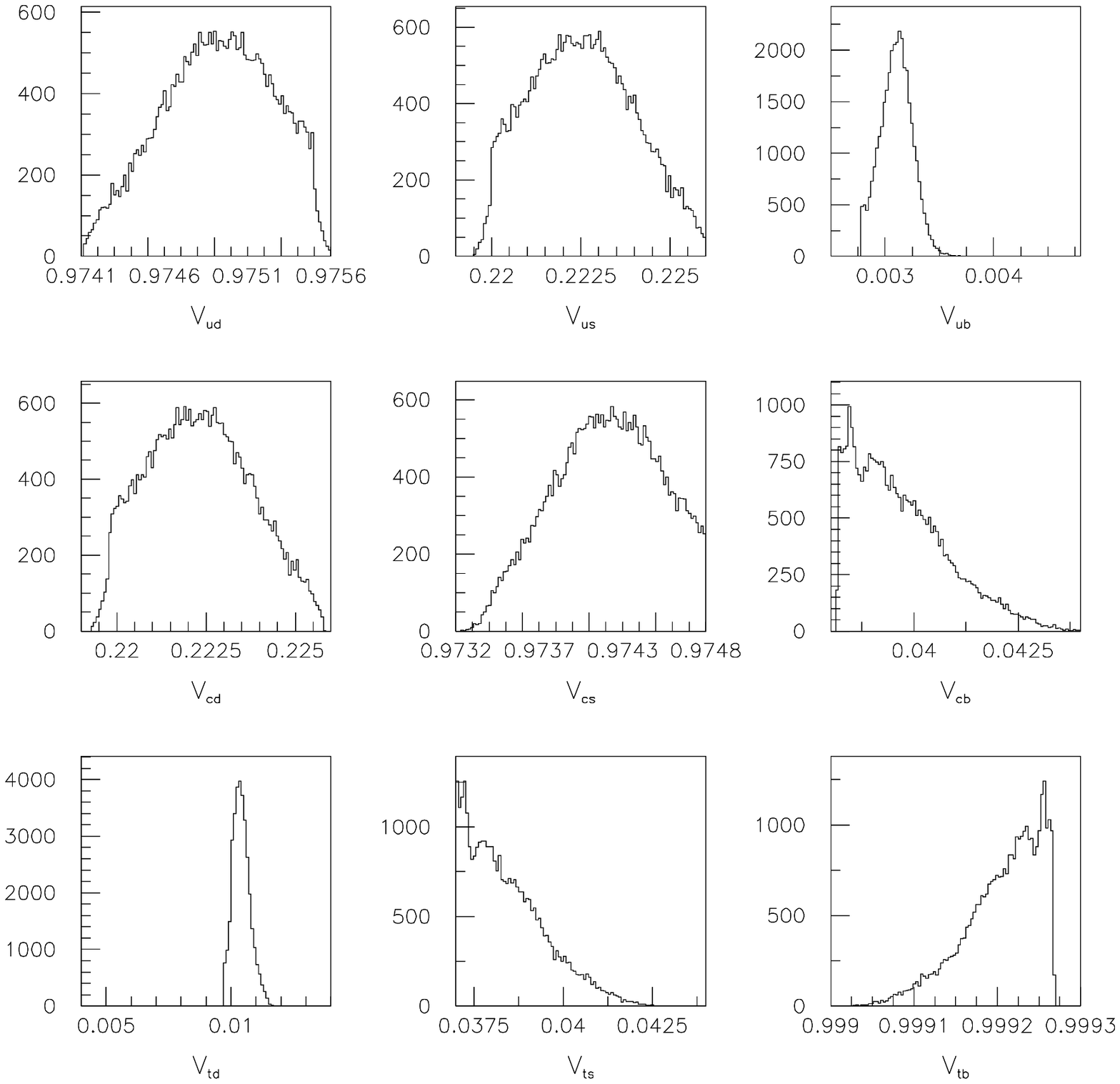,width=0.65\textwidth}
        \end{center}
        \vspace{0cm}
        \caption{Solutions for the absolute values of the CKM matrix
             elements in the case with $Q$, $U$ and $D$ localized
              at n=1. The range for each element is given by the edges of the
                 corresponding window. \label{Figure_3}}
        \end{figure}

\begin{figure}
        \begin{center}
        \epsfig{figure=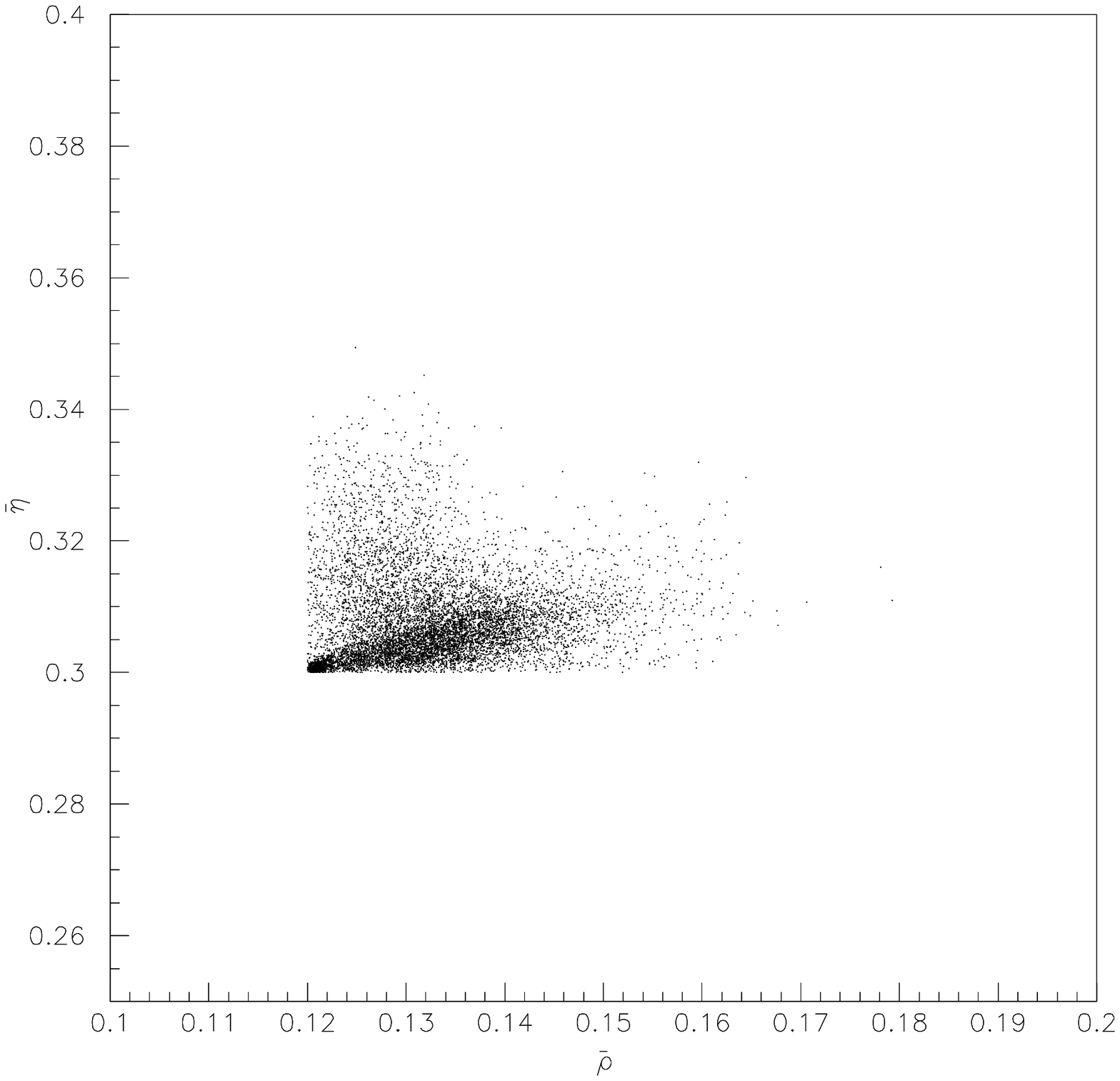,width=0.45\textwidth}
        \end{center}
        \vspace{0cm}
     \caption{Solutions for $\bar{\rho}$ and $\bar{\eta}$ in the 
               case with $Q$, $U$ and $D$ localized at n=1. \label{Figure_4}}
        \end{figure}

In the following we present the characteristically important
numerical results for the four cases out the eight mentioned
above, for which we were able to find solutions. The cases are
referred to in the above order. For each case we give one
particular, but typical, numerical complete set of the 20 defining
parameters (Table II), the quark mass matrices and quark mass
spectra, the CKM matrix and the CP parameters. Complex phases are
measured in radiant, and $N=10$ for all cases. The masses are
given in GeV and are evaluated at the $M_Z$ scale. For the sake of
visualization, we also present graphically the comprehensive
solutions of the quark wave function profiles in the theory
space (Fig. \ref{Figure_1}), the mass spectrum (Fig.
\ref{Figure_2}), the CKM matrix (Fig. \ref{Figure_3}) and the
$\bar{\rho}$-$\bar{\eta}$ CP parameters (Fig. \ref{Figure_4}) for
the case of all fields $Q,U$ and $D$ localized at the same site
$n=1$.
\bi \It Case 1: $Q$, $U$ and $D$ localized in $n=1$. \ei

\begin{equation}
M_{u}^{(QUD1)}= 78.4\,GeV\,\left(\begin{array}{ccc}
0.925\,e^{-0.558\,i}  & 0.923\,e^{-0.501\,i}  & 0.951\,e^{-0.570\,i}  \\
0.027\,e^{2.009\,i}  & 0.027\,e^{2.046\,i}  & 0.029\, e^{2.006\,i}\\
0.948\,e^{0.306\,i}  & 0.942\,e^{0.367\,i}  & 0.973\,e^{0.280\,i}
\end{array}\right) \, ,
\label{matrixupQUD1}
\end{equation}
\begin{equation}
m_{u}^{(QUD1)} = 0.0021 \,GeV\,,\,\,\,\,
m_{c}^{(QUD1)}=0.702 \,GeV\,,\,\,\,\,
m_{t}^{(QUD1)}=181.1 \,GeV\, , \label{eigenupQUD1}
\end{equation}

\begin{equation}
M_{d\,}^{(QUD1)}=1.35 \, GeV \,\left(\begin{array}{ccc}
0.909\,e^{1.490\,i} & 0.782\,e^{-2.072\,i}   & 0.960\,e^{0.992\,i}  \\
0.030\,e^{-2.649\,i}  & 0.048\,e^{0.930\,i}   & 0.032\,e^{-3.025\,i}  \\
0.848\,e^{2.339\,i}  & 0.799\,e^{-1.353\,i} & 0.918\,e^{1.838\,i}
\end{array}\right) \, ,
\label{matrixdownQUD1}
\end{equation}
\begin{equation}
m_{d}^{(QUD1)} = 0.0045 \,GeV\,,\,\,\,\,
m_{s}^{(QUD1)}=0.106 \,GeV\,,\,\,\,\,
m_{b}^{(QUD1)}=2.89 \,GeV\, . \label{eigendownQUD1}
\end{equation}

In Eqs. (\ref{matrixupQUD1}), (\ref{matrixdownQUD1}) the mass
matrices are written in a form that better shows deviations from
the democratic structure. In Eq. (\ref{CKMQUD1}) we give the
expression for the CKM matrix, in Eq. (\ref{rhoetaQUD1}) the
values for the CP parameters $\bar{\rho}$ and $\bar{\eta}$.

\begin{equation}
V_{CKM}^{(QUD1)}=\left(\begin{array}{ccc}
 0.975 - 0.009 \,i & -0.151 - 0160 \,i & -0.001 - 0.003 \,i \\
 0.015 + 0.219\, i & -0.669 + 0.709 \,i & 0.029 + 0.024 \,i \\
 0.003 - 0.009 \,i &  0.029 - 0.023 \,i & 0.670 + 0.742 \,i
\end{array}\right) \, ,
\label{CKMQUD1}
\end{equation}

\begin{equation}
\bar{\rho}^{(QUD1)} = 0.12\, , \,\,\,\,\,
\bar{\eta}^{(QUD1)} = 0.30\,  ,\label{rhoetaQUD1}
\end{equation}

\noindent with $\bar{\rho}$ and $\bar{\eta}$ defined as
\begin{equation}
\bar{\rho} = Re(V_{ud}V_{ub}^*V_{cd}^*V_{cb})/|V_{cd}V_{cb}^*|^2
\, ,
\end{equation}
\begin{equation}
\bar{\eta} = Im(V_{ud}V_{ub}^*V_{cd}^*V_{cb})/|V_{cd}V_{cb}^*|^2
\, .
\end{equation}

\bi \It Case 2: $Q$ and $U$ localized in $n=1$,
$D$ localized in $n=N$. \ei

\begin{equation}
M_{u}^{(QU1DN)}= 66.6\,GeV\,\left(\begin{array}{ccc}
0.918\,e^{0.039\,i}  & 0.609\,e^{-0.590\,i}  & 0.924\,e^{0.135\,i}  \\
0.941\,e^{0.038\,i}  & 0.637\,e^{-0.601\,i}  & 0.946\, e^{0.132\,i}\\
0.930\,e^{0.058\,i}  & 0.622\,e^{-0.585\,i}  & 0.935\,e^{0.154\,i}
\end{array}\right) \, ,
\label{matrixupQU1DN}
\end{equation}
\begin{equation}
m_{u}^{(QU1DN)} = 0.0020 \,GeV\,,\,\,\,\,
m_{c}^{(QU1DN)}=0.687 \,GeV\,,\,\,\,\,
m_{t}^{(QU1DN)}=168.3 \,GeV\, , \label{eigenupQU1DN}
\end{equation}

\begin{equation}
M_{d}^{(QU1DN)}=23.2 \, GeV \,\left(\begin{array}{ccc}
0.041\,e^{1.959\,i} & 0.045\,e^{-0.025\,i}   & 0.043\,e^{2.526\,i}  \\
0.037\,e^{1.854\,i}  & 0.042\,e^{0.003\,i}   & 0.042\,e^{2.570\,i}  \\
0.038\,e^{1.982\,i}  & 0.043\,e^{0.047\,i} & 0.043\,e^{2.605\,i}
\end{array}\right) \, ,
\label{matrixdownQU1DN}
\end{equation}
\begin{equation}
m_{d}^{(QU1DN)} = 0.0045 \,GeV\,,\,\,\,\,
m_{s}^{(QU1DN)}=0.084 \,GeV\,,\,\,\,\,
m_{b}^{(QU1DN)}=2.90 \,GeV\, , \label{eigendownQU1DN}
\end{equation}

\begin{equation}
V_{CKM}^{(QU1DN)}=\left(\begin{array}{ccc}
 0.975 + 0.029 \,i & -0.097 + 0.197 \,i & 0.001 + 0.003 \,i \\
 -0.168 - 0.141\, i & -0.880 + 0.420 \,i & 0.039 - 0.011 \,i \\
 0.003 + 0.010 \,i &  0.039 - 0.007 \,i & 0.999 + 0.011 \,i
\end{array}\right) \, ,
\label{CKMQU1DN}
\end{equation}

\begin{equation}
\bar{\rho}^{(QU1DN)} = 0.19\, , \,\,\,\,\,
\bar{\eta}^{(QU1DN)} = 0.33\,  . \label{rhoetaQU1DN}
\end{equation}

\bi \It Case 5: $Q$, $U$ and $D$ localized in
$n=N$. \ei

\begin{equation}
M_{u}^{(QUDN)}= 78.4\,GeV\,\left(\begin{array}{ccc}
0.887\,e^{0.494\,i}  & 0.881\,e^{0.478\,i}  & 0.913\,e^{0.577\,i}  \\
0.038\,e^{-2.066\,i}  & 0.038\,e^{-2.070\,i}  & 0.041\, e^{-2.039\,i}\\
0.895\,e^{-0.410\,i}  & 0.877\,e^{-0.429\,i}  &
0.929\,e^{-0.316\,i}
\end{array}\right) \, ,
\label{matrixupQUDN}
\end{equation}
\begin{equation}
m_{u}^{(QUDN)} = 0.0022 \,GeV\,,\,\,\,\,
m_{c}^{(QUDN)}=0.674 \,GeV\,,\,\,\,\,
m_{t}^{(QUDN)}=172.6 \,GeV\, , \label{eigenupQUDN}
\end{equation}

\begin{equation}
M_{d}^{(QUDN)}=1.37 \, GeV \,\left(\begin{array}{ccc}
0.895\,e^{-1.619\,i} & 0.776\,e^{1.668\,i}   & 0.943\,e^{-1.622\,i}  \\
0.058\,e^{2.456\,i}  & 0.040\,e^{-1.459\,i}   & 0.063\,e^{2.339\,i}  \\
0.835\,e^{-2.473\,i}  & 0.832\,e^{0.871\,i} & 0.893\,e^{-2.468\,i}
\end{array}\right) \, ,
\label{matrixdownQUDN}
\end{equation}
\begin{equation}
m_{d}^{(QUDN)} = 0.0049 \,GeV\,,\,\,\,\,
m_{s}^{(QUDN)}=0.106 \,GeV\,,\,\,\,\,
m_{b}^{(QUDN)}=2.90 \,GeV\, , \label{eigendownQUDN}
\end{equation}

\begin{equation}
V_{CKM}^{(QUDN)}=\left(\begin{array}{ccc}
 0.974 + 0.042 \,i & -0.046 + 0.220 \,i & 0.003 - 0.003 \,i \\
 0.134 - 0.180\, i & -0.676 - 0.701 \,i & 0.020 - 0.033 \,i \\
 -0.010 + 0.006 \,i &  0.022 + 0.030 \,i & 0.646 - 0.762 \,i
\end{array}\right) \, ,
\label{CKMQUDN}
\end{equation}

\begin{equation}
\bar{\rho}^{(QUDN)} = 0.31\, , \,\,\,\,\,
\bar{\eta}^{(QUDN)} = 0.30\,  . \label{rhoetaQUDN}
\end{equation}

\bi \It Case 6: $D$ localized in $n=1$, $Q$ and
$U$ localized in $n=N$. \ei

\begin{equation}
M_{u}^{(D1QUN)}= 71.2\,GeV\,\left(\begin{array}{ccc}
0.675\,e^{-1.829\,i}  & 0.824\,e^{-0.198\,i}  & 0.837\,e^{-0.732\,i}  \\
0.706\,e^{-1.783\,i}  & 0.856\,e^{-0.173\,i}  & 0.868\, e^{-0.698\,i}\\
0.671\,e^{-1.837\,i}  & 0.822\,e^{-0.207\,i}  &
0.834\,e^{-0.741\,i}
\end{array}\right) \, ,
\label{matrixupD1QUN}
\end{equation}
\begin{equation}
m_{u}^{(D1QUN)} = 0.0026 \,GeV\,,\,\,\,\,
m_{c}^{(D1QUN)}=0.725 \,GeV\,,\,\,\,\,
m_{t}^{(D1QUN)}=169.2 \,GeV\, , \label{eigenupD1QUN}
\end{equation}

\begin{equation}
M_{d}^{(D1QUN)}=26.5 \, GeV \,\left(\begin{array}{ccc}
0.026\,e^{0.996\,i} & 0.039\,e^{-3.019\,i}   & 0.044\,e^{-1.678\,i}  \\
0.027\,e^{0.868\,i}  & 0.037\,e^{-3.017\,i}   & 0.042\,e^{-1.655\,i}  \\
0.025\,e^{0.957\,i}  & 0.039\,e^{-3.050\,i} & 0.044\,e^{-1.710\,i}
\end{array}\right) \, ,
\label{matrixdownD1QUN}
\end{equation}
\begin{equation}
m_{d}^{(D1QUN)} = 0.0044 \,GeV\,,\,\,\,\,
m_{s}^{(D1QUN)}=0.088 \,GeV\,,\,\,\,\,
m_{b}^{(D1QUN)}=2.91 \,GeV\, , \label{eigendownD1QUN}
\end{equation}

\begin{equation}
V_{CKM}^{(D1QUN)}=\left(\begin{array}{ccc}
 -0.972 - 0.075 \,i & -0.069 + 0.213 \,i & 0.001 + 0.004 \,i \\
 -0.050 - 0.218\, i & 0.974 + 0.013 \,i & -0.039 - 0.016 \,i \\
 -0.004 - 0.012 \,i &  0.038 - 0.014 \,i & 0.998 + 0.044 \,i
\end{array}\right) \, ,
\label{CKMD1QUN}
\end{equation}

\begin{equation}
\bar{\rho}^{(D1QUN)} = 0.26\, , \,\,\,\,\,
\bar{\eta}^{(D1QUN)} = 0.38\,  . \label{rhoetaD1QUN}
\end{equation}
We are now ready for comments on the presented numerical solutions.
\section{Concluding comments}
In this paper we have reconstructed the observed complex mixing of
quark flavors, starting with the product group $\prod_{n=1}^N
[SU(2)\times U(1)]_n$ at a higher energy scale. The deconstruction
of this product group into the electroweak gauge group can indeed
provide all necessary components to generate such mixing.

We have built a specific models with 20 parameters to fit
the quark mass spectrum and the CP phase. However, the
numerical fit is found only for the ``preferred'' configurations
where fermion fields $Q$ and $U$ are localized at the same
position in the theory space. Arguably, this is because the
ratio $\kappa_U/\kappa_D$ of Yukawa couplings can be responsible
only for the difference in the overall scale of up and down-quark
masses, while the more hierarchical internal mass spectrum of the
up-quark sector (compared to that of the down-quark sector) would
still require a higher degree of overlapping.

As far as the structure of mass matrices is concerned, the deviation
from democracy is moderate. In all the cases, the mass matrices assume a
hierarchy with two rows (or two columns) having similar absolute
value matrix elements, with the third row (or third column) having
different values, but still similar along that row (or that
column). A quite close mass matrix structure was found in
\cite{HSS}, but in a different approach.

We did not perform a study of the dependence on the number of
deconstruction subgroups $N$. We expect anyway that the fitting
would be more feasible for larger $N$ as the wave functions and
their overlaps then can be tuned more smoothly. In the other
direction, the constraint from flavor changing neutral current that
sets an upper limit on the length of extra dimension in the split
fermion scenario (see e.g. \cite{DPQ}) is also expected to set
an upper limit on the ratio $N/V$ (between $N$ and the VEV of link
field) in the deconstruction theory. We however leave a more
careful analysis of these and other relevant phenomenological
issues for future publications.
\acknowledgments{P.Q.H and N-K.T are supported in part by the
U.S. Department of Energy under Grant No. DE-A505-89ER40518.
A.S. is supported by grant NSC
93-2811-M-002-047. N-K.T. also acknowledges the Dissertation Year Fellowship
from UVA Graduate School of Arts and Sciences.}

\begin{appendix}

\section{Fermion zero mode in dimensional deconstruction}
In this appendix we will work out the general expression of zero
eigenstate of the  matrix of the type (\ref{M^2_Q}). This mode
plays a special role because it will be identified with the SM
chiral fermions. To simplify the writing, here we denote this zero
eigenstate generally as $\{x_1,x_2,\ldots,x_{N-1}\}$ while in
Section III we will restore all omitted scripts $Q,U,D,i,j$.
\subsection{Zero-mode localization at the end-point n=1}
The equation set determining the zero eigenstate (\ref{M^2_Q}) is
\beqa
\label{1} &&|V|^2 x_1 - MV x_2 + V^2 x_3 =0 \Leftrightarrow
V^{*} x_1- M x_2 + V x_3=0\\
\label{2}
&&-MV^{*} x_1 + (M^2+|V|^2)x_2 -2MV x_3 + V^2 x_4 =0 \\
\label{3}
&&{V^{*}}^2 x_1 -2MV^{*} x_2+ (M^2+2|V|^2)x_3 -2MV x_4 + V^2 x_5 =0 \\
&&\ldots \nonu\\
\label{N-3} &&{V^{*}}^2 x_{N-5} -2MV^{*} x_{N-4}+
(M^2+2|V|^2)x_{N-3} -2MV x_{N-2}
+ V^2 x_{N-1} =0 \\
\label{N-2} &&{V^{*}}^2 x_{N-4} -2MV^{*} x_{N-3}+
(M^2+2|V|^2)x_{N-2}
+ (V^2-2MV) x_{N-1} =0 \\
\label{N-1} &&{V^{*}}^2 x_{N-3} ({V^{*}}^2 -2MV^{*}) x_{N-2}+
[M^2-M(V+V^{*})+2|V|^2]x_{N-1} =0 \, .\eeqa

After a bit of algebra, we can equivalently transform this equation set into
\beqa
\label{1cis}
&&X_1=X_2-|\rho|^2 X_3 \\
\label{2cis}
&&X_2=X_3-|\rho|^2 X_4 \\
\nonu
&&\ldots \\
\label{N-3cis}
&&X_{N-3}=X_{N-2}-|\rho|^2 X_{N-1} \\
\label{N-2cis} &&X_{N-2}=X_{N-1}-\rho X_{N-1} \, ,\eeqa where we
have introduced new parameter and variables \be \label{rho} \rho
\equiv |\rho| e^{i\theta}
\equiv\frac{V}{M}=\frac{|V|e^{i\theta}}{M} \ee \be \label{xX}
X_n\equiv (\rho^{*})^{N-n-1} x_n \;\;\;\;\;\;\;\;\;\; (n=1,..,N-1)
\, . \ee We note that $V$ (and $\rho$) is a complex parameter in
general (see appendix B). The new simple recursion relation allows
us to analytically determine the set $\{X_1,..,X_{N-1}\}$ (and
then the zero eigenstate $\{x_1,..,x_{N-1}\}$ ) for any $\rho$
(i.e. for any real $M$ and complex $V$). After some combinatorics
\footnote{Another somewhat simpler solution of the above equation
set is presented in \cite{HTn} } we obtain for $1\leq n\leq N-3$
\beqa \label{X_n} &&X_n = \\ \nonu &=&\sum_{k=0}\left(
\begin{array}{c}
N-3-n-k \\ k
\end{array} \right)(-|\rho|^2)^{k+1}X_{N-1} +
\sum_{k=0} \left(
\begin{array}{c}
N-2-n-k \\ k
\end{array} \right)(-|\rho|^2)^{k}X_{N-2}
\\\nonu
&=&\sum_{k=0}\frac{(N-3-n-k)!}{k!(N-3-n-2k)!}
(-|\rho|^2)^{k+1}X_{N-1}
+\sum_{k=0}\frac{(N-2-n-k)!}{k!(N-2-n-2k)!} (-|\rho|^2)^{k}X_{N-2}
\, , \eeqa and for $n=N-2$ (see (\ref{N-2cis})) \be
X_{N-2}=(1-\rho)X_{N-1} \, .\ee Using the equality \be
\left(\begin{array}{c} m \\ p \end{array}\right) +
\left(\begin{array}{c} m \\ p+1 \end{array}\right) =
\left(\begin{array}{c} m+1 \\ p+1 \end{array}\right) \, ,\ee we
can rewrite (\ref{X_n}) as (with $1\leq n\leq N-3$) \be X_{n} =
\left[ \sum_{k=0} \left(\begin{array}{c} N-1-n-k \\
k\end{array}\right)
(-|\rho|^2)^{k} -\rho \sum_{k=0} \left(\begin{array}{c} N-2-n-k \\
k\end{array}\right) (-|\rho|^2)^{k}\right] X_{N-1} \, .\ee Again,
using another equality \cite{GR} \be \sinh{px} = \sinh{x}
\sum_{k=0} (-1)^k \left(\begin{array}{c} p-1-k \\
k\end{array}\right) (2\cosh{x})^{p-1-2k} \, ,\ee we obtain for
$1\leq n\leq N-3$ and for $|\rho| < \frac{1}{2}$ \be \label{X_na}
X_n=\frac{2X_{N-1}}{\sqrt{1-4|\rho|^2}} \left[|\rho|^{N-n}
\sinh{(N-n)\alpha} - \rho|\rho|^{N-1-n} \sinh{(N-1-n)\alpha}
\right] \, ,\ee with \be \cosh{\alpha}\equiv \frac{1}{2|\rho|}
\Leftrightarrow \alpha \equiv \cosh^{-1} \frac{1}{2|\rho|}
\;\;\;\;\;\;\; (\alpha > 0) \, .\ee For $|\rho| > \frac{1}{2}$,
the expression of $X_N$ is similar to (\ref{X_na}) but with
hyperbolic functions ($\sinh$ and $\cosh$) replaced respectively
by trigonometric ones ($\sin$ and $\cos$).

Finally, from (\ref{xX}) we have altogether \be \label{Newx_ns}
x_n=Ce^{-in\theta}\left[\sinh{(N-n)\alpha} - e^{i\theta}
\sinh{(N-1-n)\alpha} \right] \;\;\;\;\;\;\; (1\leq n\leq N-1) \,
,\ee where $C$ is the normalization constant determined by the
normalization equation \be \sum_{n=1}^{N-1} |x_n|^2 =1 \, .\ee We
note that this normalization is nothing other than the unitarity
condition of the rotation matrix ${\cal{U}}$ (see (\ref{dial}) and
(\ref{Ux}), (\ref{Uy})).
\subsection{Zero-mode localization at the end-point n=N}
The chiral boundary conditions (CBC) and the value of parameter
$|\rho|\equiv |V|/|M|$ are two crucial factors that determine the
localization pattern of the chiral zero-mode of fermion. For e.g.
in the previous subsection we have seen that, when $|\rho| < 1/2$,
along with CBCs $Q_{1R}=Q_{NR}=0$\, , $\phi^Q_{N-1,N} Q_{N,L} =V_Q
Q_{N-1,L}$ (\ref{CBC}) we can localize the left-handed zero mode
of $Q$ field around site $n=1$ (\ref{Newx_ns}).

On the intuitive ground, we expect that the ``mirror image'' of
(\ref{CBC}) (apart from the requirement $|\rho| < 1/2$) \be
\label{yCBC} Q_{1,R}=Q_{N,R}=0\, , \;\;\;\;\;\;\;
{\phi^Q}^{\dagger}_{1,2} Q_{1,L} = V^{*}_Q Q_{2,L} \, ,\ee would
produce a left-handed zero mode of $Q$ field localized  at $n=N$.
A similar calculation indeed confirms this localization pattern.
Specifically, if we denote ${y_i}$ $(i=2,..,N)$ the zero-mode
subject to CBCs (\ref{yCBC}), and ${x_j}$ $(j=1,..,N-1)$ subject
to CBCs (\ref{CBC}) as before, we find \be \label{relation} y_i =
x_{N+1-i}^{*} \;\;\;\;\;\;\;\;\;  (i=2,..,N)\, , \ee or even more
explicitly (see (\ref{Newx_ns})) \be
y_n=Ce^{i(N+1-n)\theta}\left[\sinh{(n-1)\alpha} - e^{-i\theta}
\sinh{(n-2)\alpha} \right] \;\;\;\;\;\;\; (2\leq n\leq N) \, .
\label{y_ns} \ee
\section{The complex-valued link field VEV from broken Wilson line}
Since the link fields $\phi_{n,n+1}$ transform non-trivially under
two different groups, we may expect its VEV to be complex in
general. It is because in this case the VEV's phase could not be
rotated away in general. The standard and rigorous method to
determine the VEV is to write down and then minimize the
corresponding potential. It turns out \cite{BLS} that there always
exist ranges of potential parameters which generate complex VEV.
In this appendix, however, we just recapitulate the complexity
nature of link field VEV from the latticized extra dimension
perception which is derived in \cite{HL} in details. Though the
approach taken in this work does not strictly stem from
latticizing the fifth dimension, this perception could still serve
as the principle illustration.

To make the connection between DD theory and its latticized ED
counterpart, we interpret the link field as a Wilson line
connecting two neighboring branes \be \label{phi} \phi_{n,n+1}
\sim \exp(\int_{na}^{(n+1)a} ig \chi_y dy) \sim \exp( iga \chi_n)
\, , \ee where $\chi_n$ essentially is the ED component of gauge
field, $g$ and $a$ are gauge coupling and lattice spacing
respectively.

Following the DD symmetry breaking $[SU(2)\times U(1)]^N
\rightarrow [SU(2)\times U(1)]$, only one linear combination
$\chi_0$ of link fields remains massless at the classical level
\be \label{chi} \chi_0 = \frac{1}{\sqrt{N}}\sum_{n=1}^{N} \chi_n
\, . \ee In the leading order with radiative correction, by
minimizing the 1-loop effective potential of $\chi_0$, one obtains
\footnote{The finiteness of 1-loop effective potential requires
the mass of fermionic tower be trigonometric function of the mode
number \cite{HL}. In the continuum limit such as in a ``$S^1/Z_2$''
compactification, the orbifold boundary conditions (\ref{CBC}) can
fulfill this requirement.} \be \langle \chi_0 \rangle = \frac{2\pi
k}{ga\sqrt{N}} \;\;\;\;\;\;\;\;\;\; (k \in N) \, . \ee From
(\ref{phi}), (\ref{chi}), one see that in the leading order the
link fields assume a uniform complex VEV \be \langle \phi_{n,n+1}
\rangle \sim \exp(\frac{i2k\pi}{N}) \, . \ee Actually, this phase
can be considered arbitrary.
\section{Wave function overlap in theory space}
In this appendix we present the analytical expressions of
zero-mode wave function overlaps in the  theory space, from
which follow the mass matrix elements $M^{u,d}_{ij}$
(\ref{overlapU}), (\ref{overlapD}). These expressions in turn were
compiled using the exact solutions (\ref{Newx_ns}), (\ref{y_ns}) for
the wave functions. In what follows we use XX to denote the overlap of two
wave functions localized at the same site n=1, and XY the overlap
of the first wave function localized at n=1 and the second at n=N.
All other overlap configurations can be easily found by
virtue of relation (\ref{relation}).

It follows from Eqs. (\ref{overlapU}), (\ref{overlapD}) that
\beqa
\nonu
XX=&&\left( \sum_{n=1}^{N-1}x_{n}^{(1)*} x_{n}^{(2)} \right)
+  x_{N-1}^{(1)*}x_{N-1}^{(2)} \\\nonu
=&&\frac{C_1C_2}{4}   \left(
\frac{e^{(N-1)(i\theta_1-i\theta_2-\alpha_1-\alpha_2)}-1}{1-
e^{-(i\theta_1-i\theta_2-\alpha_1-\alpha_2)}}e^{N(\alpha_1+\alpha_2)}
(1-e^{-i\theta_1-\alpha_1})(1-e^{i\theta_2-\alpha_2})  \right. \\ \nonu
-&&
\frac{e^{(N-1)(i\theta_1-i\theta_2-\alpha_1+\alpha_2)}-1}{1-
e^{-(i\theta_1-i\theta_2-\alpha_1+\alpha_2)}}e^{N(\alpha_1-\alpha_2)}
(1-e^{-i\theta_1-\alpha_1})(1-e^{i\theta_2+\alpha_2}) \\ \nonu
-&&
\frac{e^{(N-1)(i\theta_1-i\theta_2+\alpha_1-\alpha_2)}-1}{1-
e^{-(i\theta_1-i\theta_2+\alpha_1-\alpha_2)}}e^{N(-\alpha_1+\alpha_2)}
(1-e^{-i\theta_1+\alpha_1})(1-e^{i\theta_2-\alpha_2})  \\ \nonu
+&&
\left.  \frac{e^{(N-1)(i\theta_1-i\theta_2+\alpha_1+\alpha_2)}-1}{1-
e^{-(i\theta_1-i\theta_2+\alpha_1+\alpha_2)}}e^{-N(\alpha_1+\alpha_2)}
(1-e^{-i\theta_1+\alpha_1})(1-e^{i\theta_2+\alpha_2})
 \right) \\
\label{xx20} +&& C_1 C_2
e^{i(N-1)(\theta_1-\theta_2)}\sinh{\alpha_1} \sinh{\alpha_2} \, ,
\linebreak
\\
\nonu
XY=&&
x_{1}^{(1)*} y_{2}^{(2)}
+ \left(\sum_{n=2}^{N-1} x_{n}^{(1)*}y_{n}^{(2)} \right)
+ x_{N-1}^{(1)*} y_{N}^{(2)} \\\nonu
=&&
C_1 C_2  e^{i\theta_1-2\theta_2} \left[
\sinh{(N-1)\alpha_1-e^{-i\theta_1}}\sinh{(N-2)\alpha_1}\right]\sinh{\alpha_2}
\\\nonu
+&&
\frac{C_1C_2}{4}   \left(
-\frac{e^{(N-2)(i\theta_1-i\theta_2-\alpha_1-\alpha_2)}-1}{1-
e^{-(i\theta_1-i\theta_2-\alpha_1-\alpha_2)}}
e^{i\theta_1-i\theta_2-\alpha_1-\alpha_2} e^{N\alpha_1+\alpha_2}
(1-e^{-i\theta_1-\alpha_1})(1-e^{-i\theta_2+\alpha_2})  \right. \\ \nonu
+&&
\frac{e^{(N-2)(i\theta_1-i\theta_2-\alpha_1+\alpha_2)}-1}{1-
e^{-(i\theta_1-i\theta_2-\alpha_1+\alpha_2)}}
e^{i\theta_1-i\theta_2-\alpha_1+\alpha_2}   e^{N\alpha_1-\alpha_2}
(1-e^{-i\theta_1-\alpha_1})(1-e^{-i\theta_2-\alpha_2}) \\ \nonu
+&&
\frac{e^{(N-2)(i\theta_1-i\theta_2+\alpha_1-\alpha_2)}-1}{1-
e^{-(i\theta_1-i\theta_2+\alpha_1-\alpha_2)}}
e^{i\theta_1-i\theta_2+\alpha_1-\alpha_2}  e^{-N\alpha_1+\alpha_2}
(1-e^{-i\theta_1+\alpha_1})(1-e^{-i\theta_2+\alpha_2})  \\ \nonu
-&&
\left.  \frac{e^{(N-2)(i\theta_1-i\theta_2+\alpha_1+\alpha_2)}-1}{1-
e^{-(i\theta_1-i\theta_2+\alpha_1+\alpha_2)}}
e^{i\theta_1-i\theta_2+\alpha_1+\alpha_2}  e^{-N\alpha_1-\alpha_2}
(1-e^{-i\theta_1+\alpha_1})(1-e^{-i\theta_2-\alpha_2})
 \right) \\
+&& C_1 C_2  e^{i(N-1)\theta_1-N\theta_2}\sinh{\alpha_1} \left[
\sinh{(N-1)\alpha_2-e^{-i\theta_2}}\sinh{(N-2)\alpha_2}\right] \,.
\label{xy20}
\eeqa
where $C_1, C_2$ are the normalization factors,
which are determined also from the overlap of the respective wave
function with itself.
\newpage
\section{Numerical tables}
\begin{table}[!h]
\caption{Central values and uncertainties for the masses of the 6
quarks evaluated at $M_Z$, for the two ratios $m_u/m_d$ and
$m_s/m_d$, for the absolute values of the CKM matrix elements and
the CP parameters $\bar{\rho}, \bar{\eta}$}
\begin{center}
\begin{ruledtabular}
\begin{tabular}{ccc}
$x_i$ & $<x_i>$ & $|x^{max}_i-x^{min}_i|/2$ \\ \hline \\
$m_u$ & $2.33 \times 10^{-3}$ & $0.45 \times 10^{-3}$ \\
$m_c$ & $0.685 $ & $0.061$ \\
$m_t$ & $181$ & $13$ \\
$m_d$ & $4.69 \times 10^{-3}$ & $0.66 \times 10^{-3}$ \\
$m_s$ & $0.0934 $ & $0.0130$ \\
$m_b$ & $3.00$ & $0.11$ \\
$m_u/m_d$ & $0.497$ & $0.119$ \\
$m_s/m_d$ & $19.9$ & $3.9$ \\
$|V_{ud}|$ & $0.97485$  & $0.00075$ \\
$|V_{us}|$ & $0.2225$  & $0.0035$ \\
$|V_{ub}|$ & $0.00365$  & $0.0115$ \\
$|V_{cd}|$ & $0.2225$  & $0.0035$ \\
$|V_{cs}|$ & $0.9740$  & $0.0008$ \\
$|V_{cb}|$ & $0.041$  & $0.003$ \\
$|V_{td}|$ & $0.009$  & $0.005$ \\
$|V_{ts}|$ & $0.0405$  & $0.0035$ \\
$|V_{tb}|$ & $0.99915$  & $0.00015$ \\
$\bar{\rho}$ & $0.22$ & $0.10$ \\
$\bar{\eta}$ & $0.35$ & $0.05$  \\
\end{tabular}
\end{ruledtabular}
\end{center}
\end{table}
\newpage
\begin{table}
\caption{20-parameter space solutions found in 4 different cases of the
model presented in Sec. IIIA (N=10 for all cases)}
\begin{center}
\begin{ruledtabular}
\begin{tabular}{ccccc}
&$(QUD1)$&$(QU1DN)$&$(QUDN)$&$(D1QUN)$
\\ \hline \\
$\alpha_{Q1}$ & $2.290$  & $0.208$  & $2.311$  & $0.215$    \\
$\alpha_{Q2}$ & $0.007$  & $0.236$  & $0.011$  & $0.251$    \\
$\alpha_{Q3}$ & $1.497$  & $0.220$  & $1.656$  & $0.213$    \\
$\alpha_{U1}$ & $0.771$  & $0.439$  & $0.623$  & $0.572$    \\
$\alpha_{U2}$ & $0.759$  & $1.900$  & $0.603$  & $0.823$    \\
$\alpha_{U3}$ & $0.927$  & $0.433$  & $0.722$  & $0.745$    \\
$\alpha_{D1}$ & $0.829$  & $0.027$  & $0.794$  & $0.541$    \\
$\alpha_{D2}$ & $0.535$  & $0.022$  & $0.471$  & $0.057$    \\
$\alpha_{D3}$ & $1.123$  & $0.029$  & $1.105$  & $0.022$    \\
$\theta_{Q1}$ & $1.002$  & $0.974$  & $1.234$  & $0.959$    \\
$\theta_{Q2}$ & $3.862$  & $0.983$  & $4.702$  & $0.958$    \\
$\theta_{Q3}$ & $1.678$  & $0.989$  & $1.999$  & $0.965$    \\
$\theta_{U1}$ & $1.163$  & $1.407$  & $1.276$  & $0.197$    \\
$\theta_{U2}$ & $1.100$  & $14.78$  & $1.246$  & $13.67$    \\
$\theta_{U3}$ & $1.247$  & $-5.290$ & $1.420$  & $-5.52$    \\
$\theta_{D1}$ & $-0.204$ & $-0.535$ & $-0.135$ & $-0.182$   \\
$\theta_{D2}$ & $3.098$  & $4.715$  & $2.809$  & $3.471$    \\
$\theta_{D3}$ & $0.086$ & $9.645$   & $-0.174$ & $9.499$    \\
${\kappa}_Uv/\sqrt{2}$ & $78.37$ & $66.63$ & $78.36$  & $71.21$    \\
${\kappa}_Dv/\sqrt{2}$ & $1.35$  & $23.24$ & $1.37$   & $26.48$    \\
\end{tabular}
\end{ruledtabular}
\end{center}
\end{table}
\newpage
\end{appendix}

\end{document}